\documentclass[prl,aps,twocolumn,floatfix,showpacs]{revtex4}
\usepackage{bm}
\usepackage{graphicx}

\newcommand{\be}{\begin{equation}}
\newcommand{\ee}{\end{equation}}
\newcommand{\bea}{\begin{eqnarray}}
\newcommand{\eea}{\end{eqnarray}}

\begin{document}
\def\tit#1#2#3#4#5{{#1} {\bf #2}, #3 (#4)}

\title{Spin-orbit induced spin-density wave in a quantum wire} 
\author{Jianmin Sun, Suhas Gangadharaiah, and Oleg A. Starykh}
\affiliation{Department of Physics, University of Utah, Salt Lake City, UT 84112}
\date{\today}

\begin{abstract}
We present analysis of the interacting quantum wire problem in the presence of
magnetic field and spin-orbit interaction. We show that
an interesting interplay of Zeeman and spin-orbit terms,
facilitated by the electron-electron interaction, results in the 
spin-density wave (SDW) state when the magnetic field and spin-orbit 
axes are {\sl orthogonal}. This strongly affects charge transport through the wire:
with SDW stabilized, single particle backscattering off an nonmagnetic impurity
becomes irrelevant. 
Sensitivity of the effect to the direction of the magnetic field
can be used for experimental verification of this proposal. 
\end{abstract}
\maketitle

{\sl Introduction}. Active current interest in devising schemes to manipulate
electron spin has led to several interesting  developments \cite{datta,streda,levitov}.
Most approaches rely on spin-orbit interaction,
which couples particle's momentum and spin,
in order to achieve this goal.
While significant progress has been made
in clarifying the role of spin-orbit (SO) interaction, mostly of the Rashba type \cite{rashba},
on the electric and spin transport by {\sl noninteracting} electrons
during the past few years \cite{halperin}, our understanding of the combined effect
of SO and electron-electron interactions is still limited \cite{samokhin00,gritsev05,tsvelik,zhenya}.

\begin{figure}[ht]
  \centering
   \includegraphics[width=2.0in]{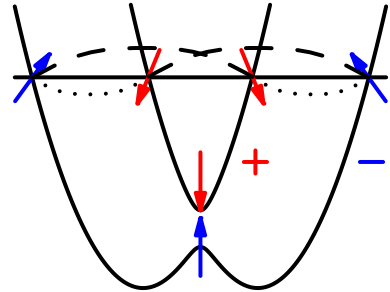}
       \caption{Occupied subbands $E_\mp$ of Eq.~\ref{eq:1}. Arrows illustrate spin polarization
    in different subbands. Dashed (dotted) lines indicate exchange (direct) Cooper scattering processes.}
  \label{fig:1}
\end{figure}

Here we study the combined effect of (Zeeman) magnetic field and
spin-orbit interaction in a single-channel {\sl interacting} quantum wire. 
This set-up allows
for the well-controlled theoretical analysis of the interplay between
broken time reversal ${\cal T}$ (by applied magnetic field) and inversion ${\cal P}$
(by spin-orbit interaction) symmetries {\sl and} electron-electron interactions.
The problem is formulated as follows. We consider a single-channel ballistic quantum wire,
corresponding to the two-terminal conductance $G_0=2e^2/h$. 
Applied magnetic field creates two spin-split subbands,
the wave functions  of which are given by the standard 
{\sl orthogonal} pair,
$\langle\uparrow|= (1,0)$ and $\langle\downarrow|= (0,1)$
(the orbital effect of the field is neglected).
It reduces spin-rotational symmetry to $U(1)$, rotations about $\hat{\sigma}_z$ axis.
Next we add {\sl weak} spin-orbit interaction $H^{(1d)}_R=\alpha_R p_y \hat{\sigma}_x$,
which is obtained by electrostatic gating  of two-dimensional electron gas with Rashba
SO interaction \cite{levitov}.
(Corrections to this form, due to the omitted
``transverse'' piece, $\alpha_R p_x \sigma_y$,
 and virtual transitions to the higher, unoccupied, subbands,
can be taken into account \cite{samokhin00} but are irrelevant for our purposes here.)
Observe that $H^{(1d)}_R$ breaks  spatial inversion ($y\to -y$) and
$U(1)$ spin symmetry, $[\hat{\sigma}_z, H^{(1d)}_R]\neq 0$.
The major consequence of this is the opening of a new, inter-subband {\sl Cooper} scattering channel
\cite{hubbard-ladder,review}. In this process, a pair of electrons
with opposite momenta in one subband is scattered (``tunnels") into a similar pair in the other
subband, see Figure~\ref{fig:1}. Note that this process {\sl requires}
spin non-conservation ({\sl i.e.} $\alpha_R \neq 0$), mentioned above,
as it scatters two electrons with (almost) ``up"-spins into a pair with 
(almost) ``down"-spins (and vice versa).
This simple observation is the key to our analysis: its derivation and consequences 
are presented below.

{\sl Technical formulation}. 
The single-particle Hamiltonian, describing the scenario outlined above, reads
\be
H_0 =  \frac{p^2}{2m} - \mu-\frac{1}{2} g\mu_B B \sigma_z + \alpha_R p \sigma_x ,
\label{eq:1}
\ee
where momentum along the wire ($y$-axis) is now denoted as $p$.
 The eigenstates
$\psi_\nu(y) = e^{i p y} \chi_\nu(p)$ $(\nu=\mp)$ are easily expressed in terms of the
momentum-dependent spinors \cite{streda,pereira}
\bea
\chi_{-}(p)= \left(
\begin{array}{c}
\cos[\gamma_p/2] \\
-\sin[\gamma_p/2] \\
\end{array}
\right) , 
\chi_{+}(p)= \left(
\begin{array}{c}
\sin[\gamma_p/2] \\
\cos[\gamma_p/2] \\
\end{array}
\right) , 
\label{eq:2}
\eea
which describe momentum-dependent orientation of electron's spin in $\hat{z}-\hat{x}$ plane.
The rotation is specified by the angle $\gamma_p=\arctan(2\alpha_R p/g\mu_B B)$.
Note that {\it left-} ({\it right-}) moving particle experiences {\it clockwise} ({\it counter-clockwise})
rotation of its spin away from ``up"-spin ($\nu= -$) and ``down"-spin ($\nu= +$) orientations
at the subband's center $p=0$, see Figure~\ref{fig:1}. The corresponding eigenvalues
$E_\mp =  \frac{p^2}{2m} - \mu  \mp \sqrt{(\alpha_R p)^2 + (g\mu_B B/2)^2}$ describe two
non-intersecting branches. The gap between them is again momentum-dependent and
is minimal at $p=0$, where it reduces to the Zeeman energy $g\mu_B B$.

We consider the situation when the Fermi energy $E_F = v_F p_F$,  
where $p_F~(v_F)$ are Fermi-momentum (velocity),
crosses both branches, as shown in Fig.~\ref{fig:1}, resulting in 
four Fermi points, $\pm p_{\mp}$ in the wire. To describe low-energy excitations of the interacting wire 
we project single-particle spin-$s$  state $\Psi_s$ ($s=\uparrow,\downarrow$) onto the
two-dimensional space spanned by $\psi_\mp$ eigenstates:
\be
\Psi_s(y) = \sum_{\nu=\mp} \langle \chi_\nu(p_\nu)| s\rangle 
 e^{ip_{\nu}y} R_{\nu} + \langle \chi_\nu(-p_\nu)| s\rangle e^{-ip_{\nu}y} L_{\nu}.
 \label{eq:3}
\ee
Operators $R_\nu$ ($L_\nu$) represent slow degrees of freedom: right- (left-) movers in the
vicinity of $+p_\nu$ ($-p_\nu$) Fermi-points of the $\nu$-th subband, respectively.
In this representation, the interaction term $H_{\text{int}} = \frac{1}{2} \sum_{s,s'}\int dy dy'
U(y-y') \Psi_s^\dagger(y) \Psi_{s'}^\dagger(y') \Psi_{s'}(y') \Psi_s(y)$ reduces to the sum
of {\it intra-} and {\it inter}-subband scattering processes \cite{review}. Keeping only
low-energy momentum-conserving ones, the {\it inter}-subband terms include,
in notations of Ref.~\onlinecite{review}, forward, exchange-backscattering and
{\it Cooper} processes. The {\it Cooper} scattering
represents two-particle ({\it pair}) tunneling between $-$ and $+$ subbands. It reads
\bea
\label{eq:4} 
&&H_{\text{C}} = \int dy ~\{U(p_{-}-p_{+})  \sin^2[(\gamma_{-}-\gamma_{+})/2] \\
&&- U(p_{-}+p_{+}) \sin^2[(\gamma_{-}+\gamma_{+})/2]\} 
(R^\dagger_{-} L^\dagger_{-} R_{+} L_{+} + \text{h.c.}) \nonumber 
\eea
Here $U(q)=\int dr U(r) e^{i q r}$ is the $q$-th Fourier component of electron interaction.
The terms inside the figure brackets in (\ref{eq:4}) represent matrix elements for two 
different Cooper scatterings - {\sl direct} and {\sl exchange}, see Fig.~\ref{fig:1}. 
$U(p_{-}-p_{+})$ describes direct scattering in which right-mover
$R_{\nu}$ in the  $\nu$-th subband scatters into right-mover $R_{-\nu}$ 
in the opposite, $-\nu$, subband, $R_{\nu} \to R_{-\nu}$,
 while its left-moving companion $L_{\nu}$ scatters into $L_{-\nu}$. 
 The other possibility, {\sl exchange} Cooper
 scattering, involving $U(p_{-}+p_{+})$, describes right and left members of the pair
scattering {\sl across}: $R_{\nu} \to L_{-\nu}$ and $L_{\nu} \to R_{-\nu}$. It is crucial to observe here that 
in addition to involving two different Fourier components of the interaction potential, these
two processes include squares of single-particle overlap integrals, 
$\sin^2[(\gamma_{-} \mp \gamma_{+})/2]$.
Relative magnitude of these is easy to understand in the limit of strong magnetic field
and weak spin-orbit splitting, $\alpha_R p_\pm/(g\mu_B B) \ll 1$, on which we concentrate now.
As discussed in the introduction, 
in this limit eigen-spinors $\chi_\mp$ {\it almost} coincide with spin $|s=\uparrow,\downarrow \rangle$
eigenstates of the Zeeman Hamiltonian.  
Weak SO term, which can be thought of as momentum-dependent magnetic field,
acting along the {\it orthogonal}, $\hat{\sigma}_x$, direction, causes spins at $p_{-}$ and
$p_{+}$ Fermi-points to {\sl tilt} by only slightly different amount, resulting in a small overlap of 
single-particle wave functions, proportional to the {\it difference} $\delta p_F = p_{-}-p_{+} = g\mu_B B/v_F$. 
At the same time, spins at, say, right $p_{-}$ and left $-p_{+}$
Fermi-points, tilt in {\sl opposite} directions, resulting in relatively large angle (and bigger overlap) 
between them, proportional to the sum $p_{-} + p_{+} = 2p_F$. This allows us to estimate
the ratio of the two amplitudes as $(U(\delta p_F)/U(2p_F)) (g\mu_B B/E_F)^2 \ll 1$ and
neglect the contribution of the {\sl direct} Cooper process in the following.

{\sl Bosonization}.
We now bosonize the problem \cite{giamarchi-book} with the help of two conjugated fields, $\varphi_\nu$ and $\theta_\nu$, 
obeying commutation relation $[\varphi_\nu(x),\theta_{\nu'}(y)]=(i/2) \delta_{\nu \nu'} (1 - \text{sign}(x-y))$.
Fermions are represented as $R_\nu = \eta_\nu \exp[i\sqrt{\pi}(\varphi_\nu - \theta_\nu)]/\sqrt{2\pi a}$ and
$L_\nu = \eta_\nu \exp[-i\sqrt{\pi}(\varphi_\nu + \theta_\nu)]/\sqrt{2\pi a}$. Klein factors $\eta_\nu$,
satisfying $\{\eta_\nu, \eta_{\nu'}\}=2\delta_{\nu \nu'}$, insure anticommutation of fermions from different subbands,
and $a\sim p_F^{-1}$ is a short-distance cutoff. We then transform to convenient symmetric, 
$\varphi_\rho=(\varphi_{-}+\varphi_{+})/\sqrt{2}$, and antisymmetric, $\varphi_\sigma=(\varphi_{-} -\varphi_{+})/\sqrt{2}$,
combinations (and similarly for $\theta_{\rho/\sigma}$), in terms of which the Hamiltonian of the problem 
decouples into two commuting ones. As indicated by notations, symmetric (antisymmetric) combinations 
in fact coincide with the standard charge (spin) ones. This is {\sl not} a generic property of the problem but rather
a convenient feature of the limit $\alpha_R p_F \ll g\mu_B B \ll E_F$ which is used in the rest of the paper.
Symmetric (charge) part $H_{\rho}$ is purely harmonic 
\be
H_\rho = \frac{1}{2}\int_y \Big(\frac{v_\rho}{K_\rho} (\partial_y \varphi_\rho)^2 + v_F (\partial_y \theta_\rho)^2\Big) ,
\label{eq:5}
\ee
with stiffness $K_\rho^{-1}=\sqrt{1+(2U(0)-U(2p_F))/\pi v_F}$.
The antisymmetric (spin) one includes nonlinear cosine term, representing Cooper process (\ref{eq:4})
\bea
&&H_\sigma = \frac{1}{2}\int_y \frac{v_\sigma}{K_\sigma} (\partial_y \varphi_\sigma)^2 + v_F (\partial_y \theta_\sigma)^2 + 
\frac{g_c}{(\pi a)^2} \cos[\sqrt{8\pi}\theta_\sigma] \nonumber\\
&&K_\sigma^{-1} = \sqrt{1-U(2p_F)/\pi v_F}, ~g_c = U(2p_F) \Big(\frac{2\alpha_R p_F}{g\mu_B B}\Big)^2 
\label{eq:6}
\eea
Renormalized velocities of these excitations follow from 
$v_{\rho,\sigma} = v_F/K_{\rho,\sigma}$. Equations (\ref{eq:5},\ref{eq:6}) 
include $H_0$ (\ref{eq:1}) as well as momentum-conserving
intra-subband (forward and backscattering) and inter-subband forward ($\propto U(0)$) interactions, 
which are encoded in the stiffnesses $K_{\rho/\sigma}$. 
Inter-subband exchange backscattering, although
momentum-conserving, is neglected because it is strictly marginal and small, of the order $\alpha_R^2$.
We have also omitted marginal correction, small in $g\mu_B B/E_F \ll 1$ factor, 
associated with weak dependence of subband velocities $v_\mp$
on magnetic field \cite{kimura} - this is the main reason for the equivalence of symmetric (antisymmetric)
modes with charge (spin) ones, mentioned above. Yet another simplification consists in replacing
$U(2p_\pm)$ by $U(2p_F)$ in expressions for $K_{\rho/\sigma}$ - this is a valid approximation for any physical $U(r)$.
Finally, we must keep
Cooper term in (\ref{eq:6}), which, inspite of having small amplitude $g_c$,
is strictly {\sl relevant} in the renormalization group (RG) sense.
Its scaling dimension is $2/K_\sigma < 2$  for repulsive interactions \cite{ao}.

Full argument in favor of Cooper term's relevancy is a bit more delicate. It has to do with irrelevant 
inter-subband direct backscattering term $\propto g_{\text{bs}} \cos[\sqrt{8\pi} \varphi_\sigma - 2 \delta p_F y]$,
omitted from (\ref{eq:6}).  Note that $K_\sigma = 1+ g_{\text{bs}}/2$.
Backscattering decays as $g_{\text{bs}}(\ell) = g_{\text{bs}}(0)/(1 + g_{\text{bs}}(0) \ell)$ until
the rescaled cutoff reaches $a(\ell) = a e^\ell \sim 1/\delta p_F$, see \cite{ao}. At that scale
$\ell^* = \ln(p_F/\delta p_F) = \ln(E_F/g\mu_B B)$ and strongly oscillating spin backscattering cosine 
disappears from the problem (``averages out") \cite{ao}. Spin stiffness $K_\sigma^* = 1 + g_{\text{bs}}(\ell^*)/2$
stops at the value {\sl above} one \cite{ao}, which implies the relevancy of the Cooper term, as 
already mentioned above.
In more detail, the Cooper coupling constant, evolution of which is described by the simple
$\partial_\ell g_c = (2-2/K_\sigma) g_c$, changes little from its 
initial value by the time scale $\ell^*$ is reached: $g_c(\ell^*) = g_c(0)[1+g_{\text{bs}}(0)\ell^*] $. 
From this point on, one is allowed to neglect $g_{\text{bs}}$ completely, and
treat the Cooper scattering term Eq.~\ref{eq:6} as the only relevant interaction.
Both $g_c$ and $K_\sigma$ grow under RG as $\ell$ is increased past $\ell^*$, and reach
strong coupling limit when $g_c(\ell) \sim v_F$ while  $K_\sigma \to 2$ \cite{gnt-book}. 

{\sl Consequences of (\ref{eq:6}).} 
The flow to strong coupling implies the change in the ground state (of spin sector) from
gapless to gapped. The resulting spin gap  can be estimated as 
$\Delta \sim [(\alpha_R p_F/g\mu_B B)^2 U(2p_F)/v_F]^{K_\sigma/(2(K_\sigma-1))}$.
This gap represents energy cost of (massive) fluctuations $\delta \theta_\sigma$ near
the semi-classical minima $\tilde{\theta}_\sigma= (m+\frac{1}{2}) \sqrt{\frac{\pi}{2}}, ~m\in Z$
of $\theta$-field. Physical meaning of these minima follows from the analysis of spin correlations.
Choosing the gauge where $\eta_\uparrow \eta_\downarrow = i$ \cite{later}, we find for the $2p_F$-components
of spin 
\bea
&& \left(
\begin{array}{c}
S^x\\
S^y\\
S^z
\end{array}
\right)_{2p_F} = -\frac{\cos[\sqrt{2\pi}\varphi_\rho + 2p_F y]}{\pi a}
\left(
\begin{array}{c}
\sin[\sqrt{2\pi}\theta_\sigma]\\
- \cos[\sqrt{2\pi}\theta_\sigma]\\
\sin[\sqrt{2\pi}\varphi_\sigma]
\end{array}
\right) 
\nonumber\\
&& \to -\frac{\cos[\sqrt{2\pi}\varphi_\rho + 2p_F y]}{\pi a}
\left(
\begin{array}{c}
\pm 1\\
0\\
0
\end{array}
\right) .
\label{eq:8}
\eea
The last line of the equation above is somewhat symbolic, with zeros representing {\sl exponentially}
decaying correlations of the corresponding spin components, $S^{y,z}$.
Here $\hat{y}$-component does not order because $\cos[\sqrt{2\pi}\tilde{\theta}_\sigma]=0$, and
$\hat{z}$-component is disordered by strong quantum fluctuations of {\sl dual} $\varphi_\sigma$ field,
as dictated by $[\varphi, \theta]$ commutation relation.
Thus ``Cooper" order found here in fact represents spin-density-wave (SDW$_x$) order  at momentum $2p_F$ 
of the $\hat{x}$-component of spin density. Observe that $S^x$ ordering is of {\sl quasi-LRO} type
as it involves free charge boson, $\varphi_\rho$. As a result, spin correlations do decay with time and
distance, but very slowly 
$\langle S^x(x) S^x(0)\rangle \sim \cos[2p_F x]~x^{-K_\rho}$. As $K_\rho < 1$ in interacting quantum wire, 
this is {\sl slower} than $x^{-1}$ decay 
typical for a one-dimensional Mott insulator - antiferromagnetic Heisenberg chain \cite{giamarchi-book}.

There is one more, very intriguing, consequence of SDW$_x$ order: suppression of $2p_F$
{\sl charge} fluctuations. Indeed, $2p_F$-component of the charge density operator reads, 
keeping subleading ($\propto \gamma_{p_F}$) contribution
\bea
\rho(y)_{2p_F} &=& - \frac{2}{\pi a} \sin[\sqrt{2\pi} \varphi_\rho + 2p_F y] \Big(
\cos[\sqrt{2\pi}\varphi_\sigma] -\nonumber\\
&&- \frac{2\alpha_R p_F}{g\mu_B B}  
\cos[\sqrt{2\pi}\theta_\sigma]\Big) \to 0
\label{eq:9}
\eea
The first term is standard, and represents {\sl intra}-subband contribution, while the second, 
involving $\theta_\sigma$, is due to the  subleading {\sl inter}-subband contribution,
which couples $\pm$ bands.
Observe that both contributions disappear in the SDW$_x$ phase
($\theta_\sigma \to \tilde{\theta}_\sigma$). Since $2p_F$-component of the charge density
describes {\sl backscattering} ($p \to -p$) of electrons by potential impurity, 
Eq.~\ref{eq:9} implies irrelevancy of the impurity in the SDW state.
The reason for this is somewhat similar to that of backscattering suppression in the spin Hall effect \cite{spinhall}:
in SDW$_x$ phase right- and left-movers within a given
subband have opposite (orthogonal)  $S^x$-components, as can be seen 
from (\ref{eq:8}) and Fig.~\ref{fig:1}, which forbids intra-subband backscattering.
(In the spin Hall case right- and left-movers
form Kramers pair and backscattering is forbidden by the ${\cal T}$ symmetry \cite{spinhall}, 
which is broken here.)
Figure~\ref{fig:1} also suggests that backscattering
between right-movers of the $+$ subband and left-movers of the $-$ one {\sl is} possible:
their $S^x$ components are parallel. Nonetheless such backscattering is still suppressed
because of  the {\sl destructive interference} of the two scattering paths. Namely,
the {\sl inter}-subband part of $2p_F$ density oscillation, bosonic form of which is given by the second term 
in (\ref{eq:9}), reads $(R^\dagger_{-} L_{+} - R^\dagger_{+} L_{-} + \text{h.c.})$ in terms of original fermions.
The crucial relative {\sl minus} sign between the two backscattering processes  
can be traced to Eq.\ref{eq:2},\ref{eq:3} and represents the noted destructive interference.
It is useful to understand this result perturbatively: 
the intra-subband piece of (\ref{eq:9}) arises from fusing $\varphi_\sigma$ from the localized impurity 
potential (first term in (\ref{eq:9})) with that in $H_R^{(1d)}$. This explains its magnitude
($\propto \alpha_R p_F/\delta p_F$) and {\sl oddness} under inversion (about the impurity site) 
${\cal P}$.  Thus, a potentially more relevant, but {\sl even} under ${\cal P}$, backscattering process
$(R^\dagger_{-} L_{+} + R^\dagger_{+} L_{-} + \text{h.c.} \sim \cos[\sqrt{2\pi} \phi_\rho] \sin[\sqrt{2\pi}\theta_\sigma])$ 
(note the relative {\sl plus} sign) can not be generated.

Although the single-particle backscattering is suppressed, the two-particle 
in general is not \cite{orignac,egger,review}. By considering fluctuations $\delta \theta_\sigma$
one indeed generates two-particle backscattering term $\propto (V^2/\Delta) \cos[\sqrt{8\pi}\varphi_\rho]$.
This spin-insensitive impurity affects finite-temperature linear conductance as
$G - 2e^2/h \propto  - (V^2/\Delta)^2 T^{4K_\rho -2}$ \cite{review}.
The correction is seen to become strong (relevant) for strongly interacting wire with 
$K_\rho < \frac{1}{2}$, when the impurity cuts off charge transport completely \cite{kane-fisher}.
This leaves us with the finite window, $\frac{1}{2} < K_\rho < 1$,
where the impurity is irrelevant. This is an interesting, and, to the best of our knowledge, {\sl new},
 addition to the Kane-Fisher result
of always relevant impurity in a single-channel repulsive Luttinger liquid \cite{kane-fisher}. Note however that 
our discussion assumes fully developed SDW$_x$ phase, and thus  implies weak disorder potential 
$V \ll \Delta$. Complete solution requires simultaneous RG analysis of  Cooper and impurity terms \cite{later}.

The correlated state can also be probed via tunneling density of states (DOS) measurements.
Skipping details, which are rather similar to the calculation of DOS in \cite{review},
we quote the result for the local DOS in SDW$_x$ state: 
$\nu(\omega) \propto \Theta(\omega - \Delta) (\omega - \Delta)^b$, where $b = (K_\rho - 1)^2/(4 K_\rho)$
and $\Theta$ denotes the step function. Naturally, DOS is zero for energies below the SDW gap,
and is found to rise smoothly ($b > 0$) just above it.  

{\it Angular stability} of the  SDW$_x$ state can be analyzed via angular dependence
of subband dispersions $E_\mp$ in Fig.~\ref{fig:1}. Indeed, suppose
that the two axes, $\vec{B}$ and SO, are not orthogonal and denote the angle between them
as $\pi/2 - \beta$. This will modify the SO term in (\ref{eq:1}) to
$\alpha_R p (\sigma_x \cos\beta  + \sigma_z \sin\beta)$.  
The eigenvalues of the modified Hamiltonian (\ref{eq:1}) now read
$E_\mp =  \frac{p^2}{2m} - \mu  \mp 
\sqrt{(\alpha_R \cos\beta p)^2 + (\frac{1}{2}g\mu_B B - \alpha_R \sin\beta p)^2}$,
and describe two subbands ($\mp$) shifted in the {\sl opposite} directions along the momentum axis.
For small $\beta$ the dispersion can be approximated as 
$E_\mp = \frac{(p \pm p_0)^2}{2m} - \mu \mp \sqrt{(\alpha_R  p)^2 + (g\mu_B B/2)^2}$.
Thus the lower ($-$) subband shifts {\sl left} and is centered around $-p_0$, while the upper ($+$) one
shifts toward positive momenta and centers around $+p_0$, where $p_0 \approx m \alpha_R \beta$.
This simple observation implies that opposite-Fermi-momenta pairs in $\pm$ subbands
acquire opposite ($\pm p_0$) {\sl center-of-mass} momenta. 
This can be pictured by shifting the bands in Fig.~\ref{fig:1} horizontally in opposite directions.
Thus two-particle Cooper tunneling processes 
 illustrated in Fig.~\ref{fig:1} become momentum-non-conserving ones.
As a result, this important scattering channel will disappear above some critical misalignment 
angle $\beta^*$ which can be estimated as follows. Cooper order is destroyed once the
misalignment cost $\propto 2 v_F p_0$ becomes comparable to the Cooper gap $\Delta$.
Estimating the latter at $K_\sigma = 2$ we find: $\beta^* \approx \alpha_R p_F U(2p_F)/(g\mu_B B)^2 \ll 1$.
This estimate shows that the found SDW$_x$ has narrow but finite region of angular stability,
and agrees fully with results of more detailed RG-based calculations in \cite{later}.
SDW$_x$ state can also be destroyed by reducing magnetic field
strength below the critical, $g \mu_B B_c \sim \alpha_R p_F$, even while maintaining the
orthogonal orientation (angle $\beta=0$). This happens due to the decrease 
of the spin stiffness $K_\sigma$ below $1$  (so that the scaling dimension of Cooper term 
(\ref{eq:6}) exceeds $2$) once the Zeeman energy becomes smaller than the spin-orbit one \cite{later}.
This weak-field region, which includes the $B=0$ limit of (\ref{eq:1}), has been studied 
previously \cite{samokhin00,gritsev05} and contains no relevant Cooper processes.

The sensitivity of the described SDW$_x$ phase to the mutual orientation and magnitude of
the magnetic and SO terms 
can be exploited in experimental searches of the novel field-induced SDW phase
of the quantum wire with spin-orbit interaction. 
It appears that lateral quantum wells at vicinal surface of gold, which 
possess spin-orbit-split and highly one-dimensional subbands,  \cite{vicinal} can serve
as a nice experimental starting point.

We would like to thank L. Balents, G. Fiete, C. Kane, D. Mattis, E. Mishchenko, L. Levitov, 
J. Orenstein, M. Raikh, and Y.-S. Wu for useful discussions and suggestions. 
O.S. and S.G. are supported by ACS PRF 43219-AC10.

\end{document}